\documentclass[aps,prl,twocolumn,showpacs]{revtex4}
\usepackage{amssymb}
\usepackage{amsmath}
\usepackage{graphicx}

\begin{document}
\title{Measurement of spin coherence using Raman scattering}
\author{Z. Sun}
\author{A. Delteil}
\author{S. Faelt}
\author{A. Imamo\u{g}lu}

\affiliation{Institute of Quantum Electronics, ETH Zurich, CH-8093
Zurich, Switzerland.}

\date{\today }

\begin{abstract}
Ramsey interferometry provides a natural way to determine the coherence time of most qubit systems. Recent experiments on quantum dots however, demonstrated that dynamical nuclear spin polarization can strongly
influence the measurement process, making it difficult to extract the $T_2^*$ coherence time using optical Ramsey pulses. Here, we demonstrate an alternative method for spin coherence measurement that is based on first-order coherence of photons generated in spin-flip Raman scattering. We
show that if a quantum emitter is driven by a weak monochromatic
laser, Raman coherence is determined exclusively by spin coherence,
allowing for a direct determination of spin $T_2^*$ time. When
combined with coherence measurements on Rayleigh scattered photons,
our technique enables us to identify coherent and incoherent
contributions to resonance fluorescence, and to minimize the latter. We verify the validity of our technique by comparing our results to those determined from Ramsey interferometry for electron and heavy-hole spins.
\end{abstract}

 \maketitle

A single electron or hole spin confined in a InGaAs self-assembled quantum dot (QD) is a promising candidate for realization of quantum information processing protocols that rely on an efficient spin-photon interface.
\cite{Gao2012, Degreeve2012, Steel2013}. For all of the proposed applications, understanding the nature of QD spin coherence using Ramsey and dynamical decoupling techniques is essential \cite{Ramsey1950, De Greve2011}. Remarkably, Ramsey interferometry implemented using optical rotation pulses in QDs is strongly influenced by dynamical nuclear spin polarization effects \cite{hogele2012, Ladd2010, carter2014}. In fact, Ramsey experiment on an electron spin shows a few non-sinusoidal oscillations before the signal vanishes completely on time scales that are a factor of $\sim 4$ shorter than the expected $T_2^*$ time.

In this Letter, we implement an alternative method to determine the spin coherence time
of a quantum emitter that is to a large extent immune to the limitations that influence Ramsey interferometry. The principal idea behind our work is the fact that first-order coherence of spin-flip Raman
scattering is determined by the coherence properties of the excitation laser
field and the spin coherence \cite{Fernandez09}. Therefore, measuring the coherence time of Raman scattered photons upon excitation with a monochromatic laser field is equivalent to a measurement of the spin dephasing time. As we show below, it is essential to carry out Raman coherence measurements at low excitation limit well below the saturation intensity in order to ensure that spin dephasing induced by Rayleigh scattering remains weak as compared to the inherent $T_2^*$ time. Moreover, dynamical nuclear spin polarization is strongly suppressed in this regime, allowing us to observe the expected Gaussian decay of the interference signal.

The experiments are based on single InGaAs QDs grown epitaxially in a p-i-n
structure. The QD layer is separated by a 35 nm tunneling barrier from the $n+$
back contact and 40 nm AlGaAs blocking barrier from the top $p+$ contact. The
p-i-n structure is placed inside a planar cavity ($Q \sim 20$), consisting of a bottom
distributed Bragg reflector (DBR) of 28 pairs and a top DBR of 2 pairs. A ZnO
solid immersion lens (SIL) mounted on the top of the sample is used to increase
the collection efficiency. The sample is held in a
bath cryostat operating at liquid He temperature. A confocal microscope
is used to excite QDs with lasers and collect scattered photons
through the same objective. The reflected laser background is suppressed to
about $10^{-6}$ by cross polarization configuration
\cite{mete2009}. The scattered photons from QDs are guided to a
superconducting single-photon detector (SSPD) and recorded by a time-correlated
single photon counting module (TCSPC). A double-$\Lambda$ system for elementary optical excitations is obtained by applying an external magnetic field perpendicular to the QD growth direction (Voigt geometry). A laser that is resonant with one of the four transitions leads to resonance fluorescence (RF), including contributions from incoherent spontaneous emission, coherent Rayleigh scattering and coherent spin-flip Raman scattering.
\begin{figure}
  \centering
  \includegraphics[width=3.5in]{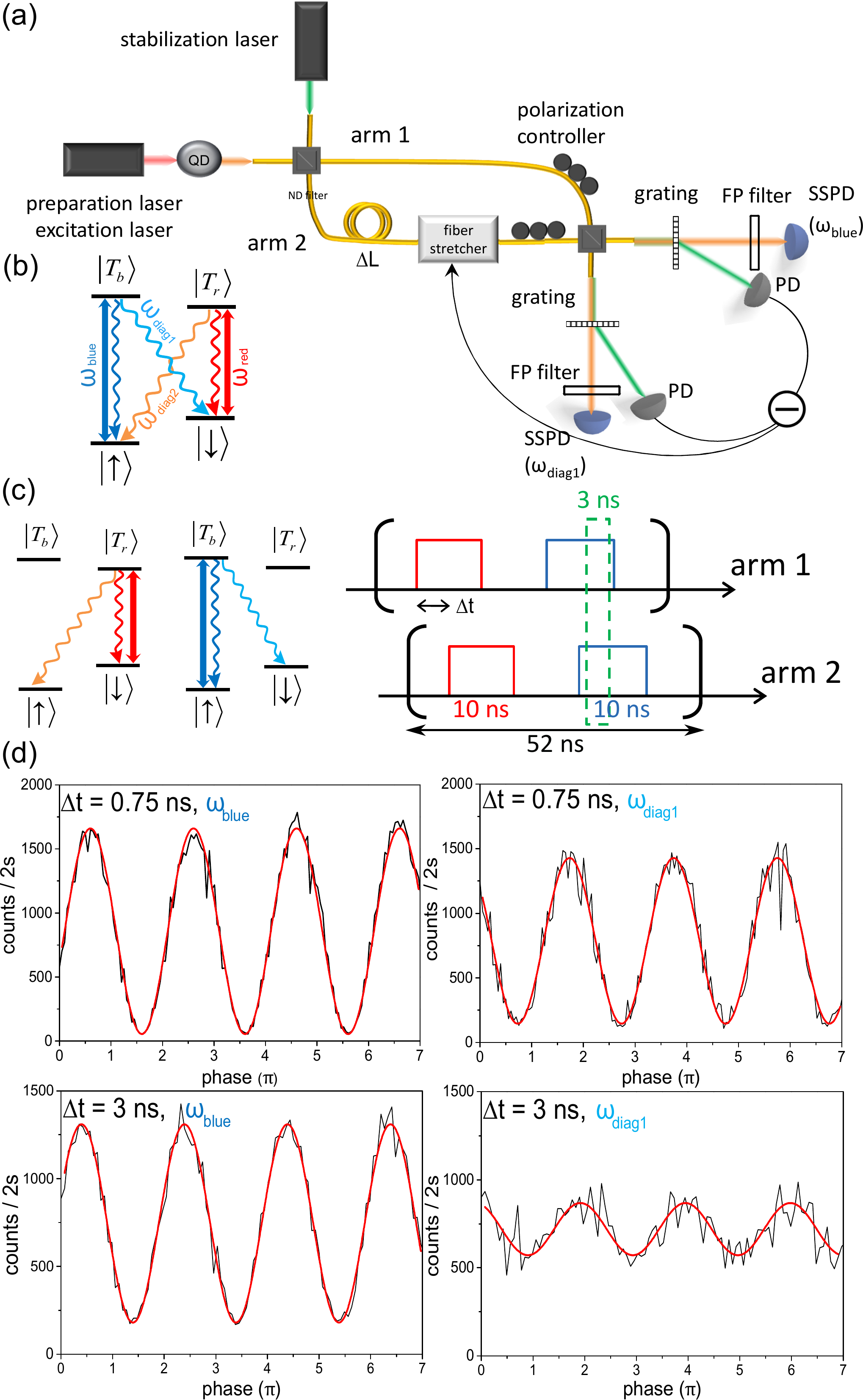}\\
  \caption{(a) Sketch of the stablized Mach-Zehnder interferometer. (b) Energy-level diagram of a single electron charged QD in Voigt geometry. (c) Pulse sequence used for the first-order coherence measurement of an electron spin and relevant transitions. Red square frame: 10~ns preparation pulse; Blue square frame: 10~ns excitation pulse; Green dashed box: 3~ns post-selected time-window. The overall repetition rate is 52~ns. (d) Count rate of SSPD as a function of phase difference, for $\Delta t = $0.75~ns (upper row) and $\Delta t = $3~ns (lower row) when filtering only $\omega_{blue}$ (left column) or $\omega_{diag1}$ (right column).}\label{1}
\end{figure}

The experiment is first performed on a single electron charged QD in Voigt geometry. Figure~1b shows the
relevant energy-level diagram in Voigt geometry \cite{Xu2007}.  A finite Zeeman splitting of the ground states and the excited states is generated by an external magnetic field $B_z$, yielding optical selection rules with four allowed transitions of identical oscillator strength.
The two ground states are identified by the orientation of the electron spin,
with $| \uparrow \rangle$ ($| \downarrow \rangle$) denoting $+1/2$ ($-1/2$)
angular momentum projection along $B_z$. We set $B_z = $4.6~T which gives to an electron Zeeman
splitting of 22~GHz and a hole Zeeman splitting of 8~GHz. The two vertical transitions (blue transition: $| \uparrow \rangle$ to blue trion state $| T_{b} \rangle$; red transition: $| \downarrow \rangle$ to red trion state $| T_{r} \rangle$) are
V-polarized, and the two diagonal transitions (diagonal transition 1: $| \downarrow \rangle$ to blue trion state $| T_{b} \rangle$; diagonal transition 2: $| \uparrow \rangle$ to red trion state $| T_{r} \rangle$) are H-polarized.
The pulse sequence used in the experiment is outlined in Figure~1c.
All the laser pulses are obtained from cw lasers by electro-optical modulators with a $10^{3}$ on$/$off ratio. The QD is first prepared in the $| \uparrow \rangle$ state by spin pumping using a 10~ns
laser pulse, termed preparation pulse, tuned on resonance with the red transition. Subsequently, a 10~ns laser pulse resonant with the blue transition, which we refer to as the excitation laser, is applied inducing a two-color photon emission: V(H)-polarized emission of center
frequency $\omega_{blue}$ ($\omega_{diag1}$) including coherent Rayleigh (Raman)
scattering and incoherent spontaneous emission.

As depicted in Figure~1a, the scattered photons are fed into one of the input ports of
stabilized Mach-Zehnder interferometer. The path length difference $\Delta L$ between the two arms leads to a time delay $\Delta t = n\Delta L/c$,
where $c$ denotes the speed of light in vacuum, $n$ denotes the refraction index of fiber. Two Fabry-P\'erot filters of 1.7~GHz linewidth select either $\omega_{blue}$ photons or $\omega_{diag1}$
photons exclusively. A TCSPC records the photon detection events, allowing
to post-select a 3~ns overlapping time-window, as shown in Figure~1c. We ensure that the 
amplitudes and polarizations of the two arms are rendered identical by introducing a variable neutral density (ND) filter in arm $2$.
To stabilize the path length difference, we use an active homodyne stabilization method
\cite{pulford2005} with an additional laser at a longer wavelength $\lambda_{0}$ such
that it can be separated from the QD photons by a transmission grating of 1500~$\ell/$mm. The photodiodes (PDs) placed at the two output ports measure the
intensity of the stabilization laser. A commercial electronic bias controller provides
a feedback signal on a fiber stretcher placed in one arm enabling to lock the path length difference to an arbitrary
value. Furthermore, it is possible to continuously change the path
length difference by scanning the stabilization laser wavelength in a quasi-static way. A
change of $d \lambda_0$ in the stabilization laser wavelength yields a change
of $d\lambda_0\Delta L/\lambda_0$ in the path length difference of the
Mach-Zehnder interferometer. As an example, Figure~1d presents the count rate of SSPD in a 3~ns post-selected time-window
for two particular time delays:  $\Delta t = 0.75$~ns (upper row) and $\Delta t = 3$~ns (lower row), when filtering only $\omega_{blue}$ (right column) or only $\omega_{diag1}$ (left column) photons.
The visibility of $\omega_{blue}$ photons is limited by the contribution from
incoherent spontaneous emission, whereas the visibility of $\omega_{diag1}$
photons is additionally reduced due to the fact that the spin-flip Raman
scattering is affected by the spin coherence.

To characterize the first-order coherence of $\omega_{blue}$ and
$\omega_{diag1}$ photons, the interference visibility is plotted
as a function of the time delay $\Delta t$ for two different powers of the excitation laser
corresponding to $P_1 = 0.1P_{sat}$, $P_2 = 0.5P_{sat}$, 
where $P_{sat} = 7.2$ nW is the saturation power for the blue transition extracted from the excitation power dependent spin pumping rate $\Gamma_{SP}$ as shown in Figure~2b (insert). In Figure~2a, the visibilities are extracted from
the interference fringes obtained by filtering either $\omega_{blue}$ or $\omega_{diag1}$ photons.
All the data have been normalized by the interference visibility
of the excitation laser in order to retain only the contribution originating from the QD scattering.

We calculate the visibility of $\omega_{blue}$ and $\omega_{diag1}$ photons using a master equation and the quantum regression theorem (QRT) (see Supplementary Information). In the simulations, we set the lifetime of $| T_{b} \rangle$ state $T_1$ to 0.76~ns which we measured from the decay of time-resolved RF excited by a short laser pulse. 
 \begin{figure}[h]
  \centering
  \includegraphics[width=3.5in]{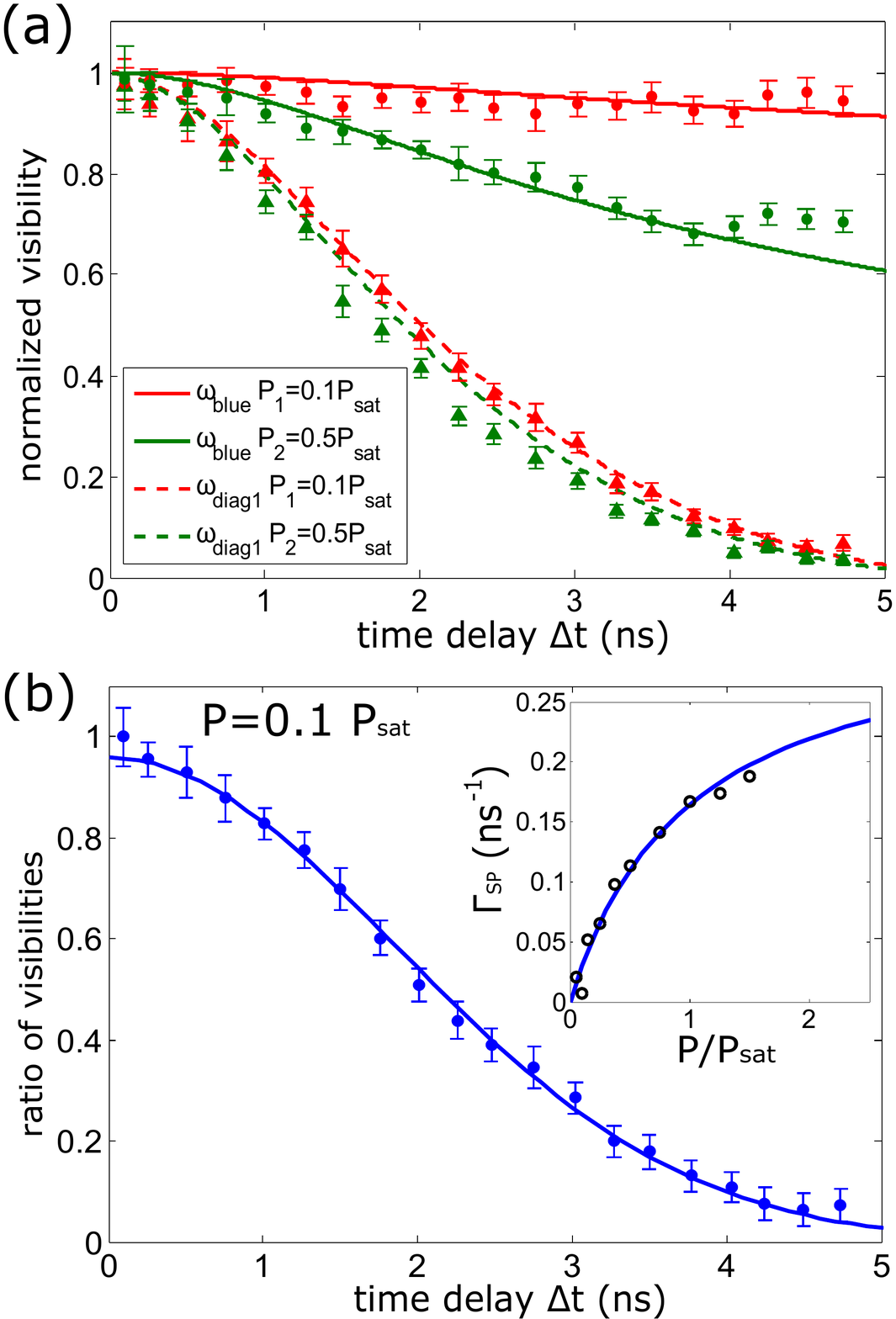}\\
  \caption{(a) The normalized visibility extracted from the interference fringes of $\omega_{blue}$ photons (solid lines and round points) and $\omega_{diag1}$ photons (dashed lines and triangular points) as a function of the time delay $\Delta t$ for two different powers of the excitation laser. In the simulation, the fitting curve is plotted using $T_1=0.76$ ns, $T_2=2T_1$, $T_2^*=2.4$ ns. (b) Electron spin: The ratio between the visibility of $\omega_{blue}$ photons and $\omega_{diag1}$ photons as a function of the time delay $\Delta t$ for the excitation laser power $P = 0.1P_{sat}$. The solid curve is a Gaussian fitting of the data. Insert: spin pumping rate as a function of the excitation laser power. The X-axis has been normalized by the saturation power of the blue transition $P_{sat} = $7.2 nW. The fitting curve is $\Gamma_{SP} = 0.5 \Omega^{2} \Gamma / (\Gamma^{2} + 2\Omega^{2})$. }\label{1}
\end{figure}
The effect of the slowly fluctuating Overhauser field is taken into account by averaging over cases with different Zeeman splitting, with a Gaussian distribution around the center value. This leads to a decay of the visibility on a time scale of  $T_2^*$. The distribution of Larmor frequencies is normal distribution with standard deviation $\sigma=\sqrt{2}/T_2^*$. The theoretical description we use is valid in the weak excitation regime where the probability to scatter more than one photon during the relevant time scale can be neglected. This implies $\Gamma_{SP}\Delta t \ll 1$.

The visibility of $\omega_{blue}$ ($\omega_{diag1}$) photons can be decomposed into a coherent Rayleigh (Raman) scattering and an incoherent spontaneous emission that vanishes on the time scale of $T_2 \leq 2T_1$, where $T_2$ denotes the coherence time of $| T_{b} \rangle$ state.
For a two-level system, the fraction of the coherent scattering is given by
\begin{equation*}
\frac{I^{coherent}}{I^{total}} = \frac{2\Gamma^2}{2\Gamma^2+{\Omega}^2} = \frac{1}{1+P/P_{sat}}
\end{equation*}
where $\Gamma = 1/T_1$ is the spontaneous emission rate of the excited state, $\Omega$ is the Rabi frequency and $P$ is the excitation laser power \cite{loudon}.  Note that at low power, there is almost no contribution from incoherent scattering for both $\omega_{blue}$ and $\omega_{diag1}$ photons. Thus the ratio between the visibility
of $\omega_{diag1}$ photons and $\omega_{blue}$ photons reveals the spin coherence allowing a direct extraction of $T_2^*$ by assuming Gaussian decay. Figure~2b is the ratio of visibilities measured with the power of the excitation laser
$P = 0.1P_{sat}$.
The $T_2^*$ of the electron spin extracted from a Gaussian fitting is 2.6~ns, in agreement with previously reported values of the electron $T_2^*$ \cite{finley2015}.

The maximum time delay $\Delta t$ at which we can still perform interference measurements is limited by the spin
pumping time, preventing us from measuring decoherence times longer than a few nanosecond. However, this limitation can be overcome by introducing a modification of the pulse sequence, which then consists of two excitation pulses of $t_0 = 3$~ns separated by $\Delta t$, matching the delay between the two arms of the interferometer, as shown in Figure~3a.
We then post-select the overlapping time-window and display the count rate. Here we illustrate this extension by measuring the hole spin dephasing time which is an order of magnitude longer than that of the electron.
The measurement is carried out on the same QD charged with a single hole in the same magnetic field. A hole is optically injected into the QD by driving the neutral exciton resonantly. By properly choosing the gate voltage, the electron tunnels out, leaving the QD with a single excess hole whose lifetime exceeds 400~ns \cite{delteil2015}. In the energy-level diagram in Figure~3a, $| \Uparrow \rangle$ ($| \Downarrow \rangle$) denotes the orientation of the heavy-hole pseudo-spin with angular momentum +3/2 ($-$3/2) projection along $B_z$. Figure~3b is the ratio of visibilities measured with an excitation laser power
$P = 0.05P_{sat}$. 
We extract $T_2^* = $ 25.7~ns from a Gaussian fitting which is consistent with previously reported values of hole coherence at high magnetic fields \cite{carter2014,greilich2011}.

The spin dephasing time of both electron and hole spin determined by the first-order coherence measurements agree with what we estimated from a Ramsey experiment on the same QD (see Supplementary Information). As mentioned earlier, the Ramsey fringes strongly deviate from a Gaussian decay of sinusoidal oscillations due to the strong hyperfine interaction. The technique presented here essentially eliminates these unwanted effects allowing an unambiguous extraction of $T_2^*$. This is rendered possible by using very low laser powers and, in the case of the electron spin, a pulse sequence that stays unchanged in the whole experiment in stark contrast with Ramsey interferometry which is influenced by delay-dependent nuclear spin polarization. Moreover, it does not need prior implementation of spin rotation and hence can be used in situations where such rotation is not applicable (e.g. to measure the quantum dot spin coherence in Faraday geometry).
\begin{figure}
  \centering
  \includegraphics[width=3.5in]{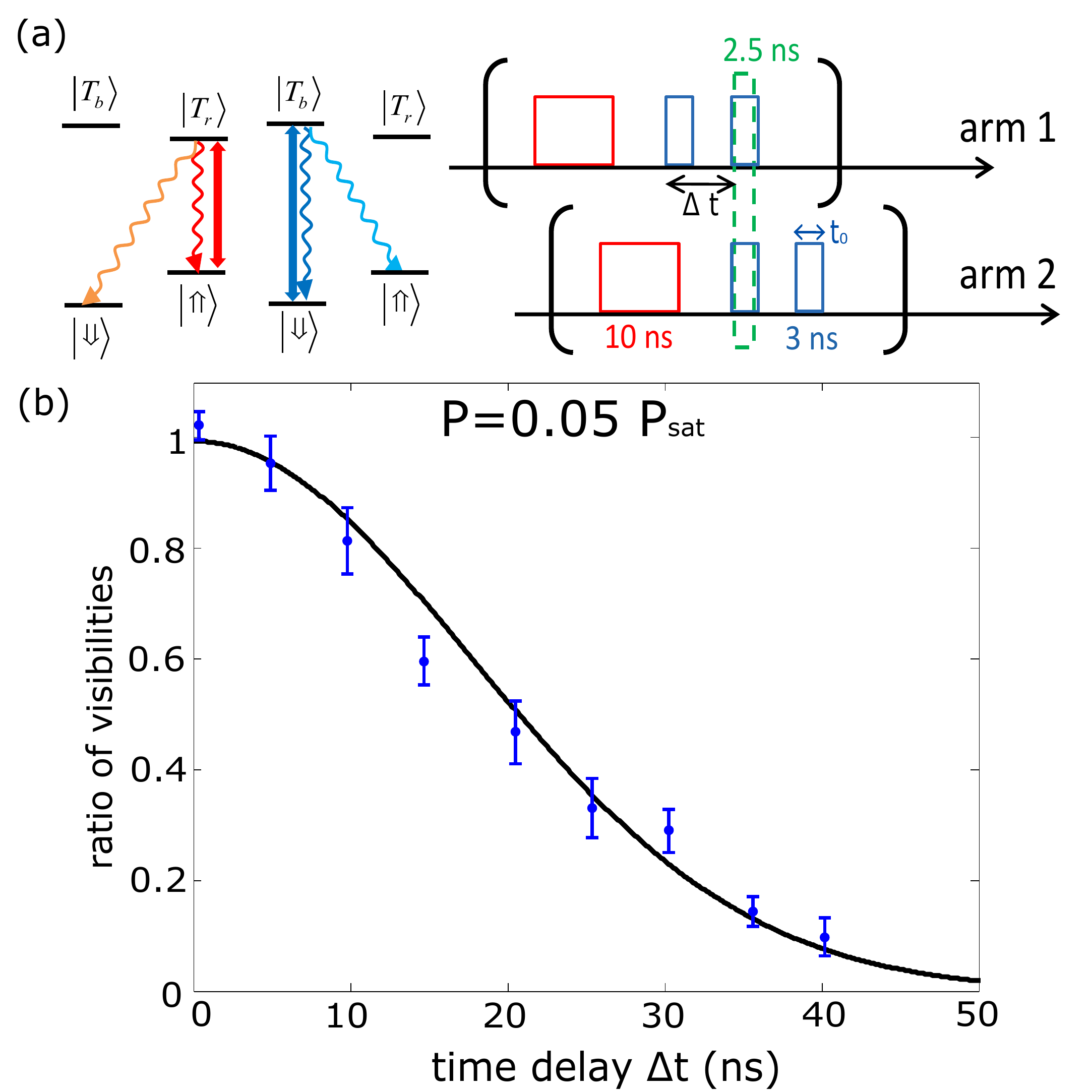}\\
  \caption{(a) Pulse sequence used for the first-order coherence measurement of a hole spin and relevant transitions. Red square frame: 10~ns preparation pulse; blue square frames: 3~ns excitation pulse; green dashed box: 2.5~ns post-selected time-window. (b) Hole spin: The ratio between the visibility of $\omega_{blue}$ photons and $\omega_{diag1}$ photons as a function of the time delay $\Delta t$ for the excitation laser power $P = 0.05P_{sat}$. The solid curve is a Gaussian fitting of the data.}\label{1}
\end{figure}

Our results demonstrate that Mach-Zehnder-type single-photon interferometry carried out on spin-flip Raman scattering can be used to measure the coherence time of a QD spin. While the use of coherent population trapping for determining spin coherence is based on the same principle \cite{imamoglu2006, warburton2009, Weiss2012}, we emphasize that the technique we present allows for a direct measurement of $T_2^*$ time that does not require the determination of the Rabi frequency associated with the driving laser field. 

This work is supported by NCCR Quantum Science and Technology (NCCR QSIT), research instrument of the Swiss National Science Foundation (SNSF) and by Swiss~NSF under Grant No. 200021-140818. The research leading to these results has received funding from the European Union Seventh Framework Programme (FP7/2007-2013) under grant agreement No.~289795. Z. S. and A. D. contributed equally to this work.


\begin{thebibliography}{99}

\bibitem{Gao2012} W.~B.~Gao, P.~Fallahi, E.~Togan, J.~Miguel-Sanchez, and A.~Imamo\u{g}lu, Nature \textbf{491}, 426 (2012).

\bibitem{Degreeve2012} K.~De Greve et al., Nature \textbf{491}, 421 (2012).

\bibitem{Steel2013} J.~R.~Schaibley, A.~P.~Burgers, G.~A.~McCracken, L.-M.~Duan, P.~R.~Berman, D.~G.~Steel, A.~S.~Bracker, D.~Gammon, and L.~J.~Sham, Phys. Rev. Lett. \textbf{110}, 167401 (2013).


\bibitem{Ramsey1950} N.~F.~Ramsey, Phys. Rev. \textbf{78}, 695 (1950).

\bibitem{De Greve2011} K.~De Greve et al., Nature Phys. \textbf{7}, 872 (2011).

\bibitem{hogele2012} A.~H\"ogele, M.~Kroner, C.~Latta, M.~Claassen, I.~Carusotto, C.~Bulutay, and A.~Imamo\u{g}lu, Phys. Rev. Lett. \textbf{108}, 197403 (2012).

\bibitem{Ladd2010} T.~D.~Ladd, D.~Press, K.~De Greve, P.~L.~McMahon, B.~Friess, C.~Schneider, M.~Kamp, S.~Ho\"fling, A.~Forchel, and Y.~Yamamoto, Phys. Rev. Lett. \textbf{105}, 107401 (2010).

\bibitem{carter2014} S.~G.~Carter, S.~E.~Economou, A.~Greilich, E.~Barnes, T.~Sweeney, A.~S.~Bracker, and D.~Gammon, Phys. Rev. B \textbf{89}, 075316 (2014).

\bibitem{Fernandez09} G.~Fernandez, T.~Volz, R.~Desbuquois, A.~Badolato, and A.~Imamo\u glu, Phys. Rev. Lett. \textbf{103}, 087406 (2009).

\bibitem{mete2009} A.~N.~Vamivakas, Y.~Zhao, C.-Y.~Lu, and M.~Atat\"ure, Nature Phys. \textbf{5}, 198 (2009).

\bibitem{Xu2007} X.~Xu, Y.~Wu, B.~Sun, Q.~Huang, J.~Cheng, D.~G.~Steel, A.~S.~Bracker, D.~Gammon, C.~Emary, and L.~J.~Sham, Phys. Rev. Lett. \textbf{99}, 097401 (2007).

\bibitem{pulford2005} D.~Pulford, C.~Robillard, and E.~Huntington, Rev. Sci. Instrum. \textbf{76}, 063114 (2005).

\bibitem{loudon} R.~Loudon, \textit{The quantum theory of light}, Oxford University Press, Oxford (2000).

\bibitem{finley2015} A.~Bechtold, D.~Rauch, F.-X Li, T.~Simmet, P.-L Ardelt,
A.~Regler, K.~M\"uller, N.~A.~Sinitsyn, and J.~J.~Finley, Nature Phys. \textbf{11}, 1005 (2015).

\bibitem{delteil2015} A.~Delteil, Z.~Sun, W.~B.~Gao, E.~Togan, S.~Faelt, and A.~Imamo\u{g}lu, Nature Phys. \textbf{12}, 218 (2015).

\bibitem{greilich2011} A.~Greilich, S.~G.~Carter, D.~Kim, A.~S.~Bracker, and D.~Gammon, Nature Photon. \textbf{5}, 702 (2011).

\bibitem{imamoglu2006} A.~Imamo\u{g}lu, Phys. Stat. Sol. (b) \textbf{243}, 3725 (2006).

\bibitem{warburton2009} D.~Brunner, B.~D.~Gerardot, P.~A.~Dalgarno, G.~W\"ust, K.~Karrai, N.~G.~Stoltz, P.~M.~Petroff, and R.~J.~Warburton, Science \textbf{325}, 70 (2009).

\bibitem{Weiss2012} K.~M.~Weiss, J.~M.~Elzerman, Y.~L.~Delley, J.~Miguel-Sanchez, and A.~Imamo\u{g}lu, Phys. Rev. Lett. \textbf{109}, 107401 (2012).



\end{thebibliography}
\end{document}


\title{Supplementary Information:\\ Measurement of spin coherence using Raman scattering}
\author{Z. Sun}
\author{A. Delteil}
\author{S. Faelt}
\author{A. Imamo\u{g}lu}

\affiliation{Institute of Quantum Electronics, ETH Zurich, CH-8093
Zurich, Switzerland.}

\date{\today }

\begin{abstract}

\end{abstract}

\pacs{03.67.Lx, 73.21.La, 42.50.-p} \maketitle

\subsection*{Ramsey fringes for the electron spin}

To verify the spin dephasing time of the electron spin extracted from first-order coherence measurement, we perform Ramsey interferometry on the QD in Voigt geometry using the pulse sequence depicted in Figure S1(a). The first pulse, termed preparation/measurement pulse, on resonance with the blue transition is used to prepare the spin in $| \uparrow \rangle$ state as well as to readout the spin state. Then we apply two $\pi$/2-pulses with a time delay $\tau$ using a 4 picosecond laser pulse detuned from the blue transition by 700 GHz. Figure S1(b) shows the counts from the QD as a function of the delay time $\tau$. The sinusoidal fringes disappear after 3 periods of Larmor precession at frequency 22 GHz, while becoming distorted before vanishing in the following 2 periods. We attribute this phenomenon to dynamic nuclear spin polarization (DNSP) effects which has been observed in previous work \cite{spin-photon, teleportation, Ladd2010}. 

To recover sinusoidal fringes, we modify the pulse sequence by inserting another preparation/measurement pulse, whose frequency is tuned in resonance with the red transition, followed by two additional $\pi$/2-pulses (Figure S1(c)) \cite{Stockill2015}. As shown in the inserted figure in Figure S1(d), the limitation of usual Ramsey experiments is partially overcome, revealing oscillations persisting at $\tau >$ 2 ns, consistent with our first-order coherence measurement. However, the damping of the Ramsey fringes still deviates from a Gaussian shape. When the time delay $\tau$ is larger than 3~ns, no oscillation can be observed any more. 

\subsection*{Ramsey fringes for the hole spin}

Similarly to the measurement performed on the electron spin, we first carry out Ramsey interferometry on the hole
spin using the usual pulse sequence which is outlined in Figure S2(a). Figure S2(b) shows the counts from the QD as a function of the delay time $\tau$. The enlarged oscillation fringes are plotted in the right part of Figure S2(b), showing a non-sinusoidal shape preventing a direct estimation of the spin dephasing time.
Again a modification of pulse sequence is applied to recover the sinusoidal fringes by alternating two patterns that differ only by the wavelength of the preparation/measurement pulse within one pulse period (Figure S2(c)). 
\begin{figure} [H]
  \centering
  \includegraphics[width=3.7in]{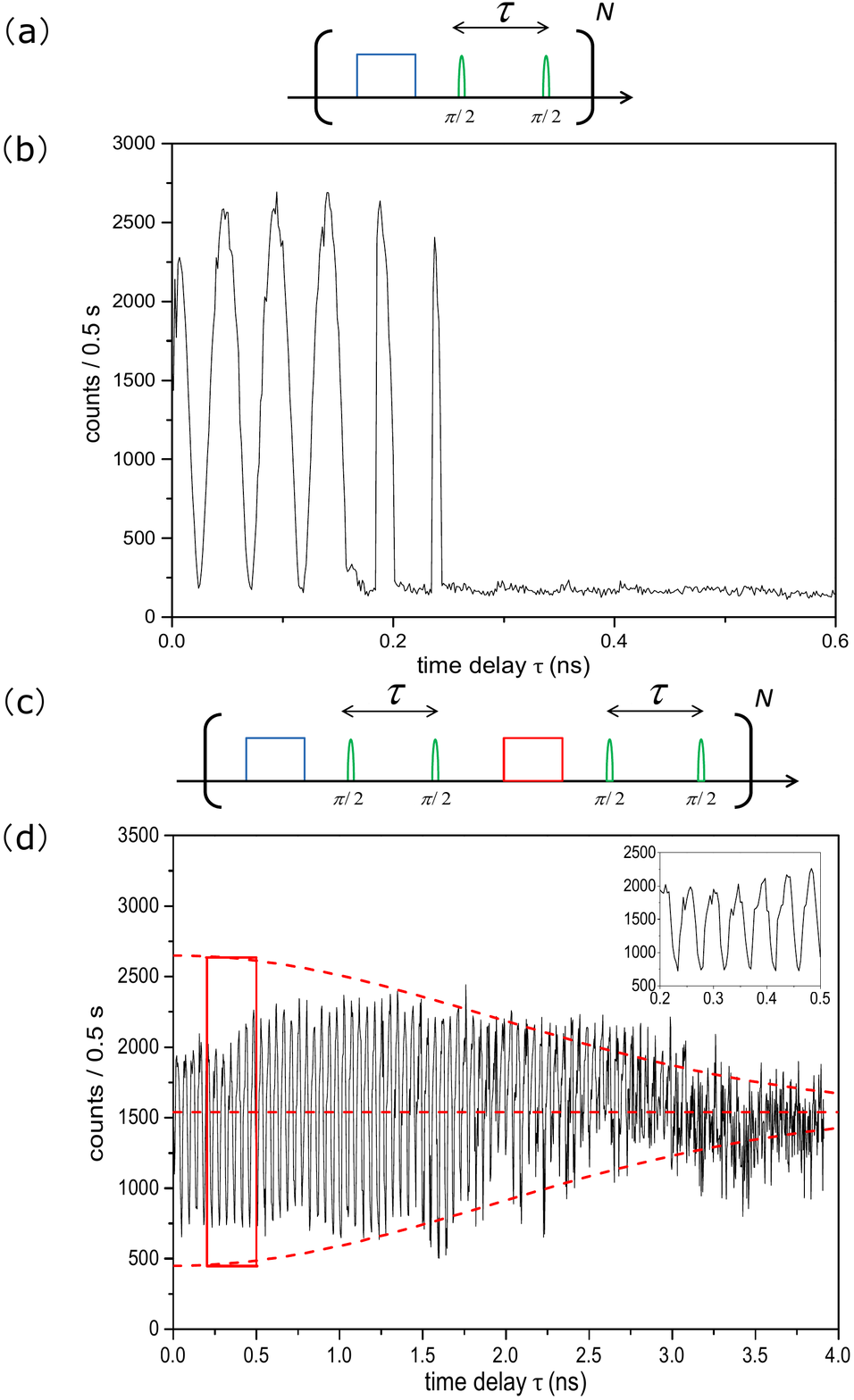}\\
\caption{ (a) Pulse sequence used for Ramsey interferometry of an electron spin. Blue square frame: 6.4~ns preparation/measurement pulse on resonance with the blue transition; Red square frame: 6.4~ns preparation/measurement pulse on resonance with the red transition; Green pulse: 4~ps laser pulse detuned by 700~GHz from the blue transition for spin rotation. (b) Ramsey fringes measured by the pulse sequence depicted in (a). Measured RF count is shown as a function of the delay time between the two $\pi$/2-pulses $\tau$. (c) Modified pulse sequence for Ramsey interferometry of an electron spin. (d) Ramsey fringes measured by the pulse sequence depicted in (c). The red dashed curves are guides for the eye using Gaussian function with $T_2^*$ extracted from first-order coherence measurement.
Insert: enlarged fringes in the time-window [0.2, 0.5] ns (red solid box).}
\end{figure}

\begin{figure}[H]
  \centering
  \includegraphics[width=3.5in]{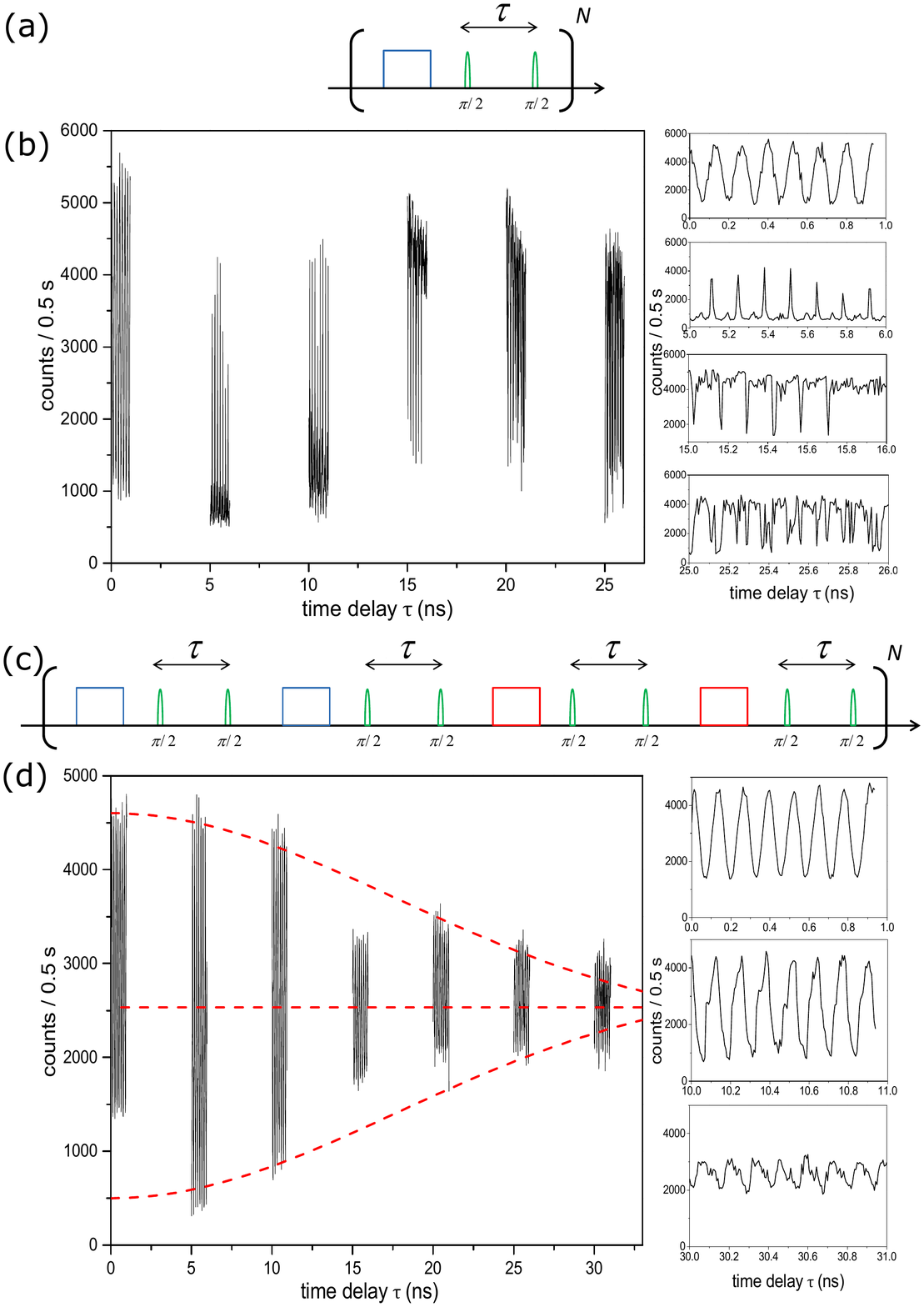}\\
\caption{ (a) Pulse sequence used for Ramsey interferometry of a hole spin. Blue square frame: 6.4~ns preparation/measurement pulse on resonance with the blue transition; Red square frame:
6.4~ns preparation/measurement pulse on resonance with the red transition; Green pulse: 4~ps laser
pulse detuned by 700~GHz from the blue transition for spin rotation. (b) Ramsey fringes measured using the pulse sequence depicted in (a). Measured RF count is shown as a function of the delay time  $\tau$ between the
two $\pi$/2-pulses. The right column displays the enlarged interference fringes. (c) Modified pulse sequence for Ramsey interferometry of a hole spin. (d) Ramsey fringes measured by the pulse sequence depicted in (c). The red dashed curves are guides for the eye using Gaussian function with $T_2^*$ extracted from first-order coherence measurement.
The right column displays the enlarged interference fringes. }
\label{figure S2}
\end{figure}
The enlarged oscillation fringes are plotted in the right part of Figure S2(d) to present the sinusoidal fringes; the envelope is non-Gaussian. Nevertheless, the interference visibility decreases clearly with increasing time delay indicating a dephasing time consistent with the value measured by first-order coherence.
 
\subsection*{Interference visibility}

Here we derive the relation between the interference visibility and the first-order correlation function. 
We investigate the first-order coherence of a $\Lambda$ system in a quasi-steady state, meaning that the quotient $\langle \sigma_{23}(t) \rangle / \langle \sigma_{33}(t) \rangle$ reaches the steady state, where $\sigma_{ij}=| i \rangle \langle j |$ ($i,j = 1, 2, 3$). The indices of the energy levels are labelled in Figure S3(a). During the measurement process, the population of the subspace $\{ | 2 \rangle, | 3 \rangle \}$ decreases due to the allowed transition $| 3 \rangle \rightarrow | 1 \rangle $ leading to an additional degradation of visibility. To obtain the decay of visibility exclusively determined by the coherence of light, we set the amplitudes of two arms identical using a neutral density filter (ND filter), equivalent to introduce an attenuation factor $\alpha$, as shown in Figure S3(b).
\begin{figure} [H]
  \centering
  \includegraphics[width=3.5in]{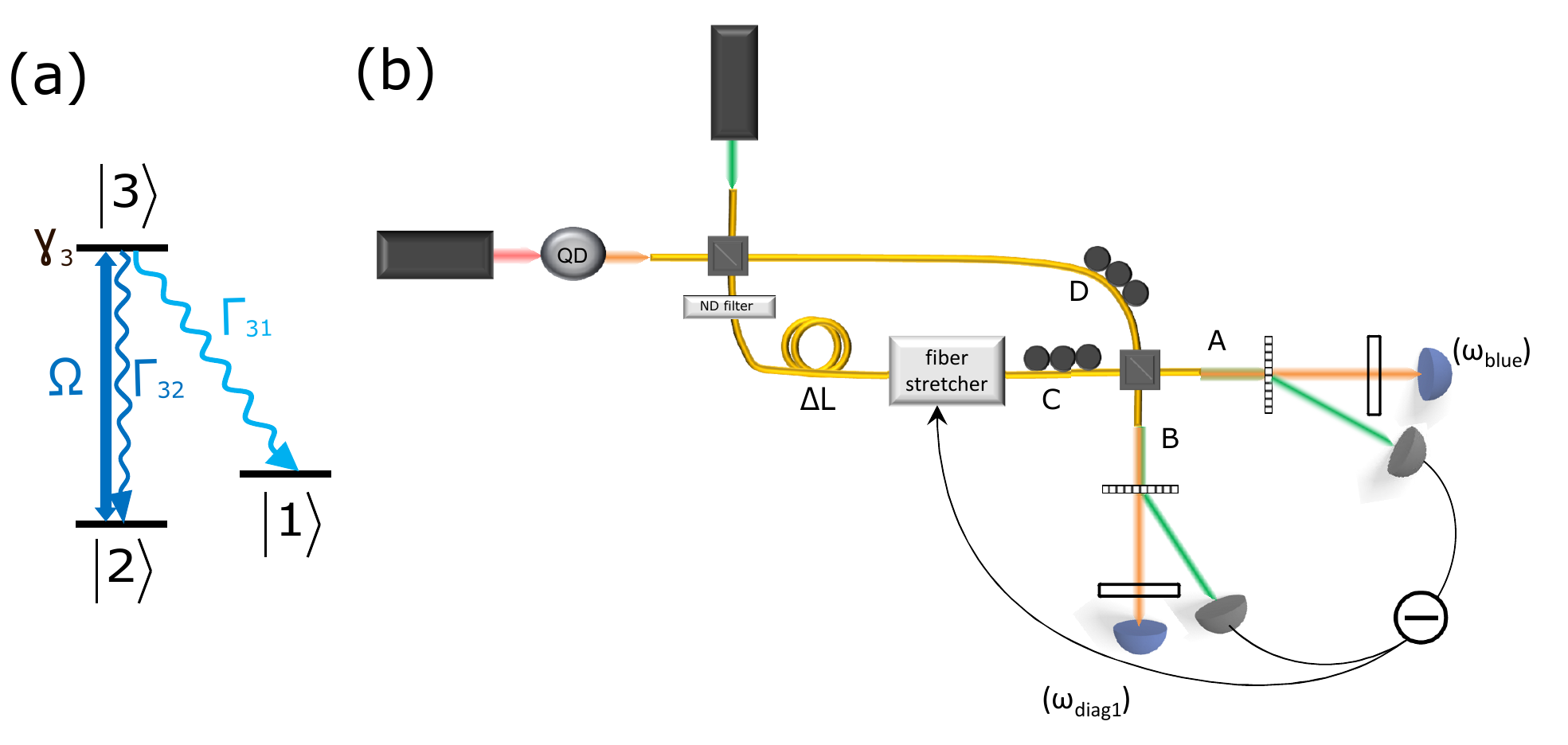}\\
\caption{ (a) Energy-level diagram of a $\Lambda$ system with all parameters. (b) Sketch of the stablized Mach-Zehnder interferometer. The output and input ports of the second beam splitter are labelled A, B, C and D respectively.} \label{figure S3}
\end{figure}

The field operator of output port $A$ or $B$ is given by
\begin{align*}
E_{A/B}^{(+)}(t+\tau)=\frac{1}{\sqrt{2}}(E_{C}^{(+)}(t+\tau) \pm E_{D}^{(+)}(t+\tau))
\\ =\frac{1}{2}(\alpha e^{i \theta}E_{QD}^{(+)}(t) \pm E_{QD}^{(+)}(t+\tau))
\end{align*}
where $E_{k}^{(+)}(t)$ is the field operator of port $k$ at time $t$, $\alpha (<1)$ is ND filter induced attenuation factor of the probability amplitude, $\theta$ is the optical phase by the interferometer. The light intensity at the port A where we select only $\omega_{blue}$ photons can be expressed as $I_{blue}=\langle E_{A}^{(-)}(t+\tau) E_{A}^{(+)}(t+\tau) \rangle$. Then the visibility of interference fringes reads 
\begin{equation*}
V_{blue}(\tau)=\frac{2 \alpha |\langle E_{QD}^{(-)}(t+\tau) E_{QD}^{(+)}(t) \rangle|}{\alpha^{2} \langle E_{QD}^{(-)}(t) E_{QD}^{(+)}(t) \rangle + \langle E_{QD}^{(-)}(t+\tau) E_{QD}^{(+)}(t+\tau) \rangle}.
\end{equation*}
The intensity of scattered light from QD decays exponentially as
\begin{equation*}
\langle E_{QD}^{(-)}(t+\tau) E_{QD}^{(+)}(t+\tau) \rangle= \langle E_{QD}^{(-)}(t) E_{QD}^{(+)}(t) \rangle e^{-\Gamma_{SP} \tau}
\end{equation*}
where $\Gamma_{SP}$ is the spin-pumping rate. To make the amplitudes of the two arms identical, the attenuation of the ND filter is set to $\alpha = exp (-\Gamma_{SP} \tau/2)$. Hence, the visibility of $\omega_{blue}$ photon is
\begin{align*}
V_{blue}(\tau)=e^{\frac{\Gamma_{SP} \tau}{2}} \frac{ |\langle E_{QD}^{(-)}(t+\tau) E_{QD}^{(+)}(t) \rangle|}{ \langle E_{QD}^{(-)}(t) E_{QD}^{(+)}(t) \rangle} 
\\ = e^{\frac{\Gamma_{SP} \tau}{2}} \frac{ |\langle \sigma_{32}(t+\tau) \sigma_{23}(t) \rangle|}{ \langle \sigma_{33}(t)\rangle}.
\end{align*}
The fluctuating Overhauser field changes the Zeeman splitting of two spin states on a time scale of $T_2^*$, such that the visibility we measured is equivalent to average over cases with different Larmor frequencies:
\begin{equation*}
V_{diag1}(\tau) =\int e^{\frac{\Gamma_{SP} \tau}{2}} \frac{ |\langle \sigma_{31}(t+\tau) \sigma_{13}(t) \rangle e^{i\omega \tau}|}{ \langle \sigma_{33}(t)\rangle} f(\omega) d\omega,
\end{equation*}
where $f(\omega)$ is the probability density of a normal distribution with mean value $\omega_{12}=22$ GHz and standard deviation $\sigma=\sqrt{2}/T_2^*$. 

\subsection*{First-order correlation of photons from a $\Lambda$ system}

In this section, we theoretically analyse a $\Lambda$ system driven by a laser resonant with one transition and calculate first-order correlation function using a master equation and the quantum regression theorem (QRT).

The interaction Hamiltonian for the system is
\begin{equation*}
H_{int}=-\frac{\Omega}{2}( \sigma_{32} + \sigma_{23}),
\end{equation*}
where $\Omega$ is the Rabi frequency of the incident laser. 
The energy level diagram of the $\Lambda$ system is plotted in Figure S3(a). The excited state spontaneously decays into two ground states with rate $\Gamma_{31}= \Gamma_{32}= \Gamma /2$, where $\Gamma$ is the spontaneous emission rate of the excited state. $\gamma_{3} = 1/T_2 - 1/(2T_1)$ denotes the pure dephasing rate of excited state $| 3 \rangle$. 

The master equation for the density matrix $\rho$ is expressed as
\begin{multline*}
\frac{d}{dt} \rho = -i[H_{int}, \rho]
\\ +\frac{\Gamma_{31}}{2}(2\sigma_{13} \rho \sigma_{31} - \sigma_{33} \rho - \rho \sigma_{33})
\\ +\frac{\Gamma_{32}}{2}(2\sigma_{23} \rho \sigma_{32} - \sigma_{33} \rho - \rho \sigma_{33})
\\ +\frac{\gamma_{3}}{2}(2\sigma_{33} \rho \sigma_{33} - \sigma_{33} \rho - \rho \sigma_{33}).
\\
\end{multline*}
The optical Bloch equations expressed in terms of the matrix elements $\rho_{ij} = \langle i | \rho | j \rangle$ takes the form
\begin{equation*}
\frac{d}{dt} \overrightarrow{\rho}= \textbf{\emph{M}}  \overrightarrow{\rho},
\end{equation*}
where $\overrightarrow{\rho} = [\rho_{11}, \rho_{12}, \rho_{13}, \rho_{21}, \rho_{22}, \rho_{23}, \rho_{31}, \rho_{32}, \rho_{33}]^{T}$. The matrix $\textbf{\emph{M}}$ reads
\begin{equation*}
\begin{bmatrix} 
0 & 0 & 0 & 0 & 0 & 0 & 0 & 0 & {\Gamma_{31}}\\ 
0 & 0 & {-\frac{i\Omega}{2}} & 0 & 0 & 0 & 0 & 0 & 0\\
0 & {-\frac{i\Omega}{2}} & {-\frac{\xi}{2}} & 0 & 0 & 0 & 0 & 0 & 0\\ 
0 & 0 & 0 & 0 & 0 & 0 & {\frac{i\Omega}{2}} & 0 & 0\\
0 & 0 & 0 & 0 & 0 & -{\frac{i\Omega}{2}} & 0 & {\frac{i\Omega}{2}} & {\Gamma_{32}}\\
0 & 0 & 0 & 0 & -{\frac{i\Omega}{2}} & {-\frac{\xi}{2}} & 0 & 0 & {\frac{i\Omega}{2}}\\
0 & 0 & 0 & {\frac{i\Omega}{2}} & 0 & 0 & {-\frac{\xi}{2}} & 0 & 0\\
0 & 0 & 0 & 0 & {\frac{i\Omega}{2}} & 0 & 0 & {-\frac{\xi}{2}} & {-\frac{i\Omega}{2}}\\
0 & 0 & 0 & 0 & 0 & {\frac{i\Omega}{2}} & 0 & {-\frac{i\Omega}{2}} & {-\Gamma}\\
\end{bmatrix}
\end{equation*}
where $\xi = \Gamma + \gamma_{3}$.

All exponential decay rates of first-order correlation can be determined by the characteristic roots of $\textbf{\emph{M}}$ (the root of characteristic polynomial $\emph{det} \left| s\textbf{\emph{I-M}} \right| = 0$): 
\begin{equation*}
s_{1}=0;
\end{equation*} 
\begin{equation*}
s_{2}\approx-\Gamma_{SP}; 
\end{equation*}
\begin{equation*}
s_{3,4}\approx-\frac{5}{8}\Gamma\pm i\Omega\quad(\Omega>\Gamma); 
\end{equation*}
\begin{equation*}
s_{5}=-\frac{\Gamma}{2}; 
\end{equation*}
\begin{equation*}
s_{6,7}=- \frac{\Gamma+\gamma_{3}}{4} + \sqrt{\left(\frac{\Gamma+\gamma_{3}}{4}\right)^{2}-\frac{\Omega^{2}}{4}};
\end{equation*}
\begin{equation*}
s_{8,9}=- \frac{\Gamma+\gamma_{3}}{4} - \sqrt{\left(\frac{\Gamma+\gamma_{3}}{4}\right)^{2}-\frac{\Omega^{2}}{4}}.
\end{equation*}
To evaluate the $\langle \sigma_{32}(t+\tau) \sigma_{23}(t) \rangle$, QRT is applied to vector $\overrightarrow{\zeta}=[\sigma_{11}(t+\tau)\sigma_{23}(t), \sigma_{12}(t+\tau)\sigma_{23}(t),\quad...\quad ,\sigma_{32}(t+\tau)\sigma_{23}(t),\sigma_{33}(t+\tau)\sigma_{23}(t)]^T$ \cite{Atac book}:
\begin{equation*}
\frac{d}{d\tau} \overrightarrow{\zeta}= \textbf{\emph{M}}  \overrightarrow{\zeta}.
\end{equation*}
This can be solved using the Laplace transform 
\begin{equation*}
s \overrightarrow{\zeta}(s)-\overrightarrow{\zeta}(0)= \textbf{\emph{M}}  \overrightarrow{\zeta}(s);
\end{equation*}
\begin{equation*}
\overrightarrow{\zeta}(s)=\frac{\emph{Adj} \left| s\textbf{\emph{I-M}} \right|^{T}}{\emph{det} \left| s\textbf{\emph{I-M}} \right|} \overrightarrow{\zeta}(0),
\end{equation*} 
where $\emph{Adj} \left| s\textbf{\emph{I-M}} \right|^{T}$  denotes the transposed adjoint of matrix $\left| s\textbf{\emph{I-M}} \right|$, $\emph{det} \left| s\textbf{\emph{I-M}} \right| = \prod\limits_{i=1}^{9}(s-s_i) $ denotes the determinant of matrix $\left| s\textbf{\emph{I-M}} \right|$. 
$\langle\sigma_{32}(t+\tau)\sigma_{23}(t)\rangle$ can be expressed using the inverse Laplace transform of matrix elements of $\emph{Adj} \left| s\textbf{\emph{I-M}} \right|^{T}$: 
\begin{multline*}
\frac{\langle\sigma_{32}(t+\tau)\sigma_{23}(t)\rangle}{\langle\sigma_{33}(t)\rangle} =
\\ \mathcal{L}^{-1}\left(\frac{\emph{Adj} \left| s\textbf{\emph{I-M}} \right|^{T}_{85}}{\emph{det} \left| s\textbf{\emph{I-M}} \right|}\right)\frac{\langle\sigma_{23}(t)\rangle}{\langle\sigma_{33}(t)\rangle}+\mathcal{L}^{-1}\left(\frac{\emph{Adj} \left| s\textbf{\emph{I-M}} \right|^{T}_{88}}{{\emph{det} \left| s\textbf{\emph{I-M}} \right|}}\right).
\end{multline*}
During the measurement, the system is in the regime where $\langle \sigma_{23}(t) \rangle / \langle \sigma_{33}(t) \rangle = i\Gamma/\Omega$ reaches a steady state. 

$\langle \sigma_{31}(t+\tau) \sigma_{13}(t) \rangle$ can also be calculated in a similar way. 
In Figure~2a in the main text, the fitting curve is plotted using $T_1=0.76$ ns, $T_2=2T_1$, $T_2^*=2.4$ ns.